
\documentclass[12pt,a4paper]{amsart}
\usepackage{amstex2latex}

\newcommand {\tto} {\longrightarrow}

\pageheight{22cm}
\pagewidth{16.5cm} \advance\oddsidemargin-2.15cm
\advance\evensidemargin-2.15cm

\define  \ad{\operatorname{ad}}
\define \adx {(\operatorname{ad} x)}
\define  \rank{\operatorname{rank}}

\define \fag  {{\goth{ag}}}
\define \fab {{\goth{ab}}}
\define \fd {{\goth{d}}}
\define \fdiff {{\goth{diff}}}
\define \fe {{\goth{e}}}
\define \ff {{\goth{f}}}
\define \fF {{\goth{F}}}
\define \fg {{\goth{g}}}
\define \fgl {{\goth{gl}}}
\define \fI {{\goth{I}}}
\define \fpsq {{\goth{psq}}}
\define \fo  {{\goth{o}}}
\define \fsp  {{\goth{sp}}}
\define \fosp  {{\goth{osp}}}
\define \fpo  {{\goth{po}}}
\define \fsl {{\goth{sl}}}
\define \fsvect {{\goth{svect}}}

\define \brk {$\newline\indent$}
\define \brki {\right.$\newline\indent$\left.}

\topmatter

\title[Defining relations]{Defining relations associated with the principal
$\fsl (2)$-subalgebras of simple Lie algebras}\endtitle

\affil Department of Mathematics, University of Stockholm \endaffil

\author Pavel Grozman, Dimitry Leites\endauthor

\address Roslagsv. 101,
Kr\"aftriket hus 6, S-104 05, Stockholm, Sweden;
mleites\@math.su.se
\endaddress

\thanks We are thankful to G.~Post for timely information about
\cite{PH} and help.  Financial support of the Swedish Institute
and NFR is gratefully acknowledged.  V. Kornyak (JINR, Dubna)
checked the generators and relations with an independent program.
\endthanks



\abstract The notion of defining relations is, clearly,
well-defined for any nilpotent Lie algebra.  Therefore, a
conventional way to present a simple Lie algebra $\fg$ is by
splitting it into the direct sum of a commutative Cartan
subalgebra and two maximal nilpotent subalgebras $\fg_{\pm}$
(positive and negative).  Though there are many (about
$(3\cdot\rank \fg)^2$) relations between the $2\cdot\rank \fg$
generators of $\fg_{\pm}$ (separately), they are neat; they are
called {\it Serre relations}.  The generators of $\fg_{\pm}$
generate $\fg$ as well.

It is possible to define the notion of relations for generators of
different type.  For instance, with the principal embeddings of
$\fsl (2)$ into $\fg$ one can associate only {\bf two} elements
that generate $\fg$; we call them {\it Jacobson's generators}.  We
explicitly describe the associated with the principle embeddings
of $\fsl (2)$ presentations of simple Lie algebras, all finite
dimensional and certain infinite dimensional ones; namely, of the
Lie algebra \lq \lq of matrices of a complex size" realized as a
subalgebra of the Lie algebra of differential operators in 1
indeterminate.

The relations obtained are rather simple, especially for
non-exceptional algebras.  In contradistinction with the
conventional presentation there are just 9 relations between
Jacobson's generators for $\fsl(\lambda)$ series and not many more
for other finite dimensional algebras.

Our results might be of interest in applications to integrable systems
(like vector-valued Liouville (or Leznov-Saveliev, or 2-dimensional
Toda) equations and KdV-type equations) based on the principal
subalgebras $\fsl (2)$.  They also indicate how to $q$-quantize the
Lie algebra of matrices of complex size.
\endabstract

\endtopmatter
\document

\specialhead Introduction\endspecialhead

This is our paper published in: Dobrushin R., Minlos R., Shubin M. and
Vershik A. (eds.)  {\it Contemporary Mathematical Physics} (F.A.
Berezin memorial volume), Amer.  Math.  Soc.  Transl.  Ser.  2, vol.
175, Amer.  Math.  Soc., Providence, RI (1996) 57--68.  We just wish
to make it more accessible.

This paper continues the description of presentations of simple Lie
superalgebras.  It is the direct continuation of \cite{LSe} and
\cite{LP}, where the case of the simplest (for computations) base is
considered and where non-Serre relations are first described, though
in a different setting.

In what follows we describe some ``natural'' generators and relations
for {\it simple finite dimensional Lie algebras} over ${\Bbb C}$.  The
answer is important in questions when it is needed to identify an
algebra given its generators and relations.  (Examples of such are
Eastbrook-Vahlquist prolongations, Drienfield's quantum algebras,
symmetries of differential equations, integrable systems, etc.).

If $\fg$ is nilpotent, the problem of its presentation has a natural
and unambiguous solution: representatives of $\fg/ [\fg, \fg]$ are
generators of $\fg$ and $H_2(\fg)$ describes relations.

On the other hand, if $\fg$ is simple, then $\fg =[\fg, \fg]$ and
there is no ``most natural'' way to select generators of $\fg$.  The
choice of generators is not unique.

Still, among algebras with the property $\fg =[\fg, \fg]$, the
simple ones are distinguished by the fact that their structure is
very well known.  By trial and error people discovered that, for
finite dimensional simple Lie algebras, there are certain ``first
among equal'' sets of generators:

1) {\it Chevalley generators} corresponding to positive and negative
simple roots;

2) a pair of generators that generate any finite dimensional simple
Lie algebra associated with the {\it principal $\fsl(2)$-subalgebra}
(considered below).

The relations associated with Chevalley generators are well-known, see
e.g., \cite{OV}.  These relations are called {\it Serre relations}.

The possibility to generate any simple finite dimensional Lie
algebra by two elements was first claimed by N.~Jacobson (an
exercise in \cite{J}); for the first (as far as we know) proof,
see \cite{BO}.  We do not know what generators Jacobson had in
mind; \cite{BO} take for them linear combinations with generic
coefficients of positive and negative root vectors, respectively;
nothing like a \lq\lq natural" choice of what we suggest to refer
to as {\it Jacobson's generators} was ever proposed: to generate a
simple algebra with only two elements is tempting but nobody did
so explicitly, yet.  To check whether the relations between these
elements are nice-looking is impossible without a computer (cf. an
implicit description in \cite{F}).  We did it with the help of a
{\it Mathematica}-based package developed by Grozman \cite{GL}. As
far as we could test, the relations for any other pair of
generators chosen in a way distinct from ours are more
complicated.

One of our aims was to decipher \cite{F}.  Certain statements from
\cite{F} are clarified (also with the help of a computer) in
\cite{PH} appeared as \cite{PH1}; we use some of these clarifications in \S 2.

In what follows we explicitly list the relations between
Jacobson's generators; well, actually, for beautification of
relations we introduce a third generator.  Throughout the paper
$\fg$ is a simple Lie algebra.

\head \S 1. The case of a finite dimensional $\fg$ \endhead

\subhead 1.1.  Principal embeddings \endsubhead There exists only
one (up to equivalence) embedding $\rho: \fsl(2)\tto \fg$ such
that $\fg$, considered as $\fsl(2)$-module, splits into $\rank
\fg$ irreducible modules.  (The reader may consider this statement
as an exercise or consult \cite{D}, \cite{LS} or \cite{OV}.)  This
embedding is called {\it principal} and, sometimes, {\it minimal}
because for other embeddings (there are plenty of them) the number
of irreducible $\fsl(2)$-modules is $>\rank \fg$.  Example: for
$\fg=\fsl(n)$, $\fsp(2n)$ or $\fo(2n+1)$, the principal embedding
is the one corresponding to the irreducible representation of
$\fsl(2)$ of dimension $n$, $2n$, $2n+1$, respectively.

For completeness, let us recall how the irreducible
$\fsl(2)$-modules with highest weight look like.  (They are all of
the form $M^{\mu}=L^{\mu}$, where $\mu\not\in {\Bbb N}$, and
$L^n$, where $n\in {\Bbb N}$, described below.)  Select the
following basis in $\fsl(2)$:
$$
X^-=\pmatrix 0&0\\ -1&0\endpmatrix, \quad H=\pmatrix 1&0\\
0&-1\endpmatrix, \quad X^+= \pmatrix 0&1\\ 0&0 \endpmatrix.
$$
The $\fsl(2)$-module $M^{\mu}$ is illustrated with a graph whose nodes
are eigenvectors $l_i$ of $H$ with the weight indicated;
$$
\dots\stackrel{ -\mu}{\longleftrightarrow} \circ
-\stackrel{-\mu+2}{\longleftrightarrow} \circ -\dots
-\stackrel{\mu-2}{\longleftrightarrow} \circ -\stackrel{
\mu}{\longleftrightarrow} \circ
$$
the edges depict the action of $X^{\pm}$ (the action of $X^+$ is
directed to the right, that of $X^-$ to the left:
$X^-l_{\mu}=l_{\mu-2}$ and
$$
X^+l_{\mu-2i}=X^+((X^-)^il_{\mu})=i(\mu-i+1)l_{\mu-2i+2};\quad
X^+(l_{\mu})=0.\tag 1.1
$$

As follows from (1.1), the module $M^n$ for $n\in {\Bbb N}$ has an
irreducible submodule isomorphic to $M^{-n-2}$; the quotient,
obviously irreducible as follows from the same (1.1), will be
denoted by $L^n$.

As the $\fsl(2)$-module corresponding to the principle embedding, a
simple finite dimensional Lie algebra $\fg$ is as follows (cf.  \cite{OV},
Table 4):

\heading Table 1.1. $\fg$ as the $\fsl(2)$-module\endheading
$$

\begin{tabular}{|c|c|}
\hline
$\fg$&the $\fsl(2)$-spectrum of $\fg=L^2\oplus L^{k_1} \oplus L^{k_2} \dots$\\
\hline
$\fsl(n)$&$L^2\oplus L^4 \oplus L^6 \dots \oplus L^{2n-2}$\\
$\fo (2n+1)$, $\fsp(2n)$&$L^2\oplus L^6 \oplus L^{10} \dots \oplus L^{4n-2}$\\
$\fo (2n)$&$L^2\oplus L^6 \oplus L^{10} \dots \oplus L^{4n-2}\quad\oplus L^{2n-2}$\\
$\fg_2$&$L^2\oplus L^{10}$\\
$\ff_4$&$L^2\oplus L^{10}\oplus L^{14}\oplus L^{22}$\\
$\fe_6$&$L^2\oplus L^{8}\oplus L^{10}\oplus L^{14}\oplus L^{16}\oplus L^{22}$\\
$\fe_7$&$L^2\oplus L^{10}\oplus L^{14}\oplus L^{18}\oplus L^{22}\oplus L^{26}\oplus
L^{34}$\\
$\fe_8$&$L^2\oplus L^{14}\oplus L^{22}\oplus L^{26}\oplus L^{34}\oplus L^{38}
\oplus L^{46}\oplus L^{58}$\\
\hline
\end{tabular}
$$

One can show that $\fg$ can be generated by two elements:
$x:=\nabla_+\in L^2=\fsl (2)$ and a lowest weight vector
$z:=l_{-r}$ from an appropriate module $L^r$ other than $L^2$ from
Table 1.1. For the role of this $L^r$ we take either $L^{k_{1}}$
if $\fg\not=\fo(2n)$ or the last module $L^{2n-2}$ in the above
table if $\fg =\fo(2n)$. (Clearly, $z$ is defined up to
proportionality; we will assume that a basis of $L^r$ is fixed and
denote $z=t\cdot l_{-r}$ for some $t\in {\Bbb C}$ that can be
fixed at will.)

The exceptional choice for $\fo(2n)$ is occasioned by the fact that by choosing
$z\in L^{r}$ for $r\neq 2n-2$ instead, we generate $\fo(2n-1)$.

We call the above $x$ and $z$, together with $u:=\nabla_-\in L^2$ for good measure, {\it
Jacobson's generators}. The presence of $u$ considerably simplifies the form of the
relations, though slightly increases their number. (One might think that taking the symmetric
to $z$ element $l_r$ will improve the relations even more but in reality just the opposite
happens.)

\subhead 1.2. Relations between Jacobson's generators\endsubhead
First, observe that if an ideal of a free Lie algebra is homogeneous
(with respect to the degrees of the generators of the algebra), then the number and
the degrees of the defining relations (i.e., the generators of the ideal) is
uniquely defined provided the relations  are homogeneous. This is obvious.

A simple Lie algebra $\fg$, however, is the quotient of a free Lie
algebra $\fF$ modulo a nonhomogeneous ideal, $\fI$, the ideal
without homogeneous generators. Therefore, we can speak about the
number and the degrees of relations only conditionally. Our
conditions is the possibility for any element $x\in\fI$ to be
expressible via the generators $g_1, ...$ of $\fI$ by a formula of
the form
$$
x=\sum c_ig_i,\text{  where $c_i\in \fF$ and $\deg c_i+\deg g_i\leq \deg x$
for all $i$.} \tag*
$$
(The degree is calculated with respect to that of the generators of $\fF$.)

Under condition $(*)$ the number and the degree of relations is uniquely defined. Now we can
explain the necessity for the extra generator $y$: without it the weight relations would have
been of very high degree.

We divide the relations between Jacobson's generators into the
types corresponding the number of occurrence of $z$ in them:

{\bf
0}. Relations in $L^2 = \fsl(2)$;

{\bf 1}. Relations coming from
$L^2 \otimes L^{k_1}$;

{\bf 2}. Relations coming from
$L^{k_{1}}\wedge L^{k_1}$;

$\pmb{\geq 3}$. Relations coming from
$L^{k_1}\wedge L^{k_1}\wedge L^{k_1}\wedge \dots$ with $\geq 3$
factors; among the latter relations we distinguish one: of type
$\pmb\infty$ --- the relation bounding the dimension. (For small
$\text{rank}\, \fg$, the  relation of type $\infty$ can be of the
above types.)

Observe that, apart form relations of type $\infty$, the relations of type $\geq 3$ are those
of type 3 except for $\fe_7$ which satisfies stray relations of type 4 and 5.

The relations of type 0 are the well-known relations in $\fsl(2)$
$$
\pmb{0.1}. \; \; [[x, y], \, x]=2x,\quad \quad  \quad
\pmb{0.2}\; \;  [[x, y], \, y]=-2y.\tag0
$$
The relations of type 1 express that the space $L^{k_1}$ is the
$(k_1+1)$-dimensional $\fsl(2)$-module.  To simplify notations we
denote: $z_i=(\ad x)^iz$.

$$
\pmb{1.1}.\; \; [y, \, z] = 0,\quad \pmb{1.2}.  \; \; [[x, y], \, z] =
-k_1z,\quad \pmb{1.3}.\; \; z_{k_{1}+1} = 0\quad \text{ with $k_1$
from Table 1.1}.\tag1
$$

\proclaim{1.3. Theorem}. For the simple finite dimensional Lie
algebras, all the relations between Jacobson's generators are the above
relations $(0)$, $(1)$ and the relations from Table $3.1$.
\endproclaim

\head \S 2.  The Lie algebra $\fsl(\lambda)$ as a subalgebra of
$\fdiff (1)$ and $\fsl_+ (\infty)$\endhead

\subhead 2.1\endsubhead The Poincar\' e-Birkhoff-Witt theorem (PBW)
states that, as spaces,
$$
U(\fsl(2))\cong {\Bbb C}[X^-, H, X^+].
$$
We
also know that to study representations of $\fg$ is the same as to
study representations of $U(\fg)$.  Still, if we are interested in
irreducible representations, we do not need the whole of $U(\fg)$ and
can do with a smaller algebra, easier to study.

This observation is used now and again; Feigin applied it in \cite{F}.  As
deciphered in \cite{PH}, Feigin wrote, actually, that setting
$$
X^-=-{d\over d u}, \quad
H=2u{d\over d u}-(\lambda-1),  \quad X^+=u^2{d\over d
u}-(\lambda-1) u
$$
we obtain a realization of $\fsl(2)$ by differential operators.  This
realization can be extended to a morphism of associative algebras: $R:
U(\fsl(2))\longrightarrow {\Bbb C}[u, {d\over d u}]$.  The kernel of
$R$ is the ideal generated by $\Delta-\lambda ^2+1$, where
$\Delta=2(X^+X^-+X^-X^+)+H^2$.  Observe, that this morphism is not an
epimorphism, either.  Though not so easy to describe as $U(\fg)$ in
the PBW theorem, the image of this morphism turned out to be very
interesting: this is our Lie algebra of matrices of \lq\lq complex
size".

\remark{Remark} In their proof of certain statements from \cite{F}
that we will recall, \cite{PH} make use of the well-known fact that
the Casimir operator $\Delta$ acts on the irreducible $\fsl(2)$-module
$L^{\mu}$ with highest weight $\mu$ (i.e., $H\cdot l_{\mu}=\mu\cdot
l_{\mu}$ and $X^+l_{\mu}=0$) as the scalar operator of multiplication
by $\mu^2+2\mu$.  The passage from \cite{PH}'s $\lambda$ to \cite{F}'s
$\mu$ is done with the help of a shift by the weight $\rho$, the
half-sum of positive roots.  Since $\rho$ for $\fsl(2)$ can be
identified with 1, we have $\lambda-1=\mu$, and
$(\lambda-1)^2+2(\lambda-1)=\lambda^2-1$.  \endremark

\vskip 0.3 cm

Consider the Lie algebra $U(\fsl(2))_L$ associated with the
associative algebra $U(\fsl(2))$.  (We denote by the subscript ${}_L$
the functor that sends an associative algebra to the Lie algebra with
the bracket determined by the commutator.)  It is easy to see that, as
$\fsl(2)$-module, $$ U(\fsl(2))_L/(\Delta-\lambda ^2+1)\simeq L^0\oplus
L^2\oplus L^4\oplus\dots\oplus L^{2n}\oplus \dots \simeq {\Bbb
C}[R(\fsl(2))]_L\subset{\Bbb C}[x, \frac{d}{dx}]\tag2.1
$$

Observe the crucial difference between the associative algebra ${\Bbb
C}[\fsl(2)]$ generated by $\fsl(2)$ and the associative algebra ${\Bbb
C}[R(\fsl(2))]$ generated by the image of $\fsl(2)$ under $R$.  The
associated graded algebras are generated by 3 and 2 generators,
respectively.

It is not difficult to show (for details, see \cite{PH}) that the Lie
algebra
$$
U_n=U(\fsl(2))_L/(\Delta -(n^2-1))
$$ contains an ideal $I_{n}$
for $n\in {\Bbb N}\setminus\{0, 1\}$, and the quotient $U_n=/I_{n}$ is
the conventional $\fgl(n)$.  In \cite{PH} it is proved that, for $\lambda
\neq {\Bbb Z}$, the Lie algebra $U_\lambda=U(\fsl(2))_L/(\Delta-\lambda
^2+1)$ has only one ideal --- the space of constants.  This justifies
Feigin's suggestive notations
$$
\fsl(\lambda)=\fgl(\lambda)/<1>, \text{ where }\fgl(\lambda)= \cases
U_\lambda&\text{for $\lambda \not\in {\Bbb N}\setminus\{0, 1\}$}\\
U_n/I_{n}&\text{for $n\in {\Bbb N}\setminus\{0, 1\}$.}\endcases
 \tag2.2
$$
The definition directly implies that
$\fsl(-\lambda)\cong\fsl(\lambda)$, so speaking about real values of
$\lambda$ we can confine ourselves to the nonnegative values.

\subhead 2.2.  There is no analog of Cartan matrix for
$\fsl(\mu)$\endsubhead Are there {\it Chevalley generators}, i.e.,
elements $X^{\pm}_i$ of degree $\pm 1$ and $H_i$ of degree 0 (the {\it
degree} is the weight with respect to the $\fsl(2)=L^2\subset
\fsl(\mu)$) such that
$$
[X^+_i, X^-_j]= \delta_{ij}H_i, \quad [H_i, H_j]=0?
$$
The answer is {\bf NO}: $\fsl(\mu)$ is too small. We can complete it by considering
infinite sums of its elements, but the completion erases the difference between different
$\mu$'s:

\proclaim{Proposition}. The completion of $\fsl(\mu)$ generated
by Jacobson's generators {\rm (see Table 3.2)} is isomorphic to $\fdiff (1)$.
\endproclaim

\subhead 2.3\endsubhead
The invariants of the map
$$
X\longmapsto -X^{T}\quad \hbox{for}\;  \quad X\in\fgl(n).\tag2.3
$$
constitute $\fo(n)$ if $n\in 2{\Bbb N}+1$ or $\fsp(n)$ if $n\in 2{\Bbb N}$.
By analogy, Feigin defined $\fo(\lambda)$ and $\fsp(\lambda)$ as
subalgebras of $\fgl(\lambda)=\oplus_{k\geq 0} L^{2k}$ invariant with respect to the
involution
$$
X\longmapsto  \cases -X&\text{if $X\in L^{4k}$}\\
X&\text{if $X\in L^{4k+2}$},\endcases\tag2.4
$$
the analogue of (2.3).
Since $\fo(\lambda)$ and $\fsp(\lambda)$ --- the subalgebras of $\fgl(\lambda)$ singled
out by the involution (2.4) --- differ by a shift of the parameter $\lambda$, it is
natural to denote them uniformly (but so as not to confuse with the Lie superalgebras of
series $\fosp$), namely, by $\fo{/}\fsp(\lambda)$. For integer values of the parameter it
is clear that   $$
\fo{/}\fsp(\lambda) = \cases \fo(\lambda) & \text{if $\lambda \in 2{\Bbb
N}+1$}, \\
\fsp(\lambda) & \text{if $\lambda \in 2{\Bbb N}$}.\endcases
$$
In the realization of $\fsl(\lambda)$ by differential operators
the above involution is the passage to the adjoint operator;
hence, $\fo{/}\fsp(\lambda)$ is a subalgebra of $\fsl(\lambda)$
consisting of self-adjoint operators.

 \subhead 2.4. The Lie algebra
$\fsl(\lambda)$ as a subalgebra of $\fsl_+(\infty)$\endsubhead Recall that $\fsl_+(\infty)$
often denotes the Lie algebra of infinite (in one direction; index $+$ indicates that)
matrices with nonzero elements inside a (depending on the matrix) strip  along the main
diagonal and containing it. The subalgebras $\fo(\infty)$ and $\fsp(\infty)$ of
$\fsl(\infty)$ are naturally defined.

The realization 2.1 provides with an embedding
$\fsl(\lambda)\subset\fsl_+(\infty)=\lq\lq\fsl(M^{\lambda})"$, so for $\lambda\neq {\Bbb N}$
the Verma module $M^{\lambda}$ with highest weight $\mu$ is an irreducible
$\fsl(\lambda)$-module.

\proclaim{Proposition}. The completion of $\fsl(\lambda)$
(generated by the elements of degree $\pm 1$ with respect to $H\in
\fsl(\lambda)$) is isomorphic for any non-integer $\lambda$ to
$\fsl_+(\infty)=\lq\lq\fsl(M^{\lambda})"$.  \endproclaim

\subhead 2.5. The Lie algebras $\fsl(*)$ and $\fo{/}\fsp(*)$, for
$*\in{\Bbb C}{\Bbb P}^1={\Bbb
C}\cup\{*\}$\endsubhead
The \lq\lq quantization" of the relations for $\fsl(\lambda)$ and
$\fo{/}\fsp(\lambda)$ (see Table 3.2) is performed by passage to the limit as
$\lambda\longrightarrow\infty$ under the change:
$$ t\mapsto
\cases{t\over \lambda}&\text{for $\fsl(\lambda)$}\\
{t\over \lambda^2}&\text{for $\fo{/}\fsp(\lambda)$}.\endcases
$$
So the parameter $\lambda$ above can actually run over ${\Bbb C}{\Bbb P}^1={\Bbb
C}\cup\{*\}$, not just ${\Bbb C}$. In the realization with the help of
deformation, cf. 2.7 below, this is obvious. Denote the limit algebras by $\fsl(*)$
and $\fo{/}\fsp(*)$ in order to distinguish them from $\fsl(\infty)$ and
$\fo(\infty)$ or $\fsp(\infty)$ from sec. 2.4.

It is clear that $\fsl(*)$ and $\fo{/}\fsp(*)$ are subalgebras of the whole \lq\lq
plane" algebras $\fsl(\infty)$ and
$\fo(\infty)$ or $\fsp(\infty)$, it is impossible to embed  $\fsl(*)$
and $\fo{/}\fsp(*)$ into the \lq\lq quadrant"
algebra $\fsl_+(\infty)$.

\proclaim {2.6. Main Theorem}.
For Lie algebras $\fsl(\lambda)$ and $\fo{/}\fsp(\lambda)$, where $\lambda\in {\Bbb C}{\Bbb
P}^1$,  all the relations between Jacobson's generators are the relations of types $0, 1$
with $k_1$ found from Table $1.1$ and the relations from Tables in \S $3$. \endproclaim

\subhead 2.7\endsubhead (\cite{PH}).  Now, consider another
realization of $\fsl(2)$: before the factorization (2.2).  This
realization was a starting point of \cite{F}, so we give it for
completeness.  Take the Lie algebra $\fpo(2)_{ev}$ of even degree
polynomials ${\Bbb C}[q, p]_{ev}$ with respect to the Poisson bracket.
Set $X^-={1\over 2}q^2, X^+={1\over 2}p^2$, and notice that $<q, p>$
is the identity $\fsl(2)$-module.  Observe that, as $\fsl(2)$-modules,
the Lie algebras $\fpo(2)_{ev}$ and its deform --- the result of the
quantization, a subalgebra of the Lie algebra $\fdiff(1)$ of
differential operators on the line --- also have spectrum $(2.1)$.  So
it is natural to look at the deforms for various values of the
parameter of deformation; this is done in \cite{PH}.

Observe that the deforms of $\fpo(2)$ are all isomorphic for
nonzero values of the parameter of deformation, unlike the deforms
of subalgebra $\fpo(2)_{ev}$; indeed apart from the isomorphism
$\fsl(-\lambda)\cong\fsl(\lambda)$ all the deforms are
non-isomorphic.

\head \S 3.  Tables.  Jacobson's generators and relations between
them\endhead \subhead Table 3.1.  Finite dimensional
algebras\endsubhead

In what follows $E_{ij}$ are the matrix units; $X^{\pm}_i$ stand for
the conventional Chevalley generators of $\fg$.

\vskip 1mm
\noindent $\underline{\fsl(n)\text{ for }n\geq 3}$. Generators:
\roster
\item"" $x = \sum\limits_{1\leq i\leq n-1}i(n-i)E_{i, i+1}, \qquad
y = \sum\limits_{1\leq i\leq n-1}E_{i+1, i}, \qquad
z = t\sum\limits_{1\leq i\leq n-2} E_{i+2, i}$.
\endroster

Relations:
\roster
\item"{\bf 2.1}." $3[z_1, \, z_2] - 2[z, \, z_3] = 24t^2(n^2-4) y$,
\item"{\bf 3.1}." $[z, \, [z, z_1]] = 0$,
\item"{\bf 3.2}." $4 [z_3, \, [z, z_1] ] - 3 [z_2, \, [z, z_2]]
     = 576 t^2(n^2-9)z$.
\item"$\pmb{\infty=n-1}$." $(\ad z_1)^{n-2}z = 0$
\endroster

\noindent For $n=3, 4$ the degree of the last relation is lower than the
degree of some other relations, this yields a simplification:

\noindent$\underline{n = 4}$:
\roster
\item"{\bf 2.1}." $3[z_1, \, z_2] - 2[z, \, z_3] = 288 t^2 y$,
\item"{\bf 3.1}." $[z, \, [z, z_1]] = 0$,
\item"{\bf 3.2}." $[z_3, \, [z, z_1] ] = -576 t^2 z$,
\item"$\pmb{\infty=3}$." $(\ad z_1)^2 z = 0$.
\endroster

\noindent $\underline{n = 3}$: $\pmb{\infty=2}$. $[z_1, z] = 0$,
{\bf 2.1}. $[z_1, \, z_2] = 24 t^2 y$.

\noindent $\underline{\fo(2n+1)\text{ for }n\geq 3}$. Generators:
\roster
\item"" $x = n(n+1)(E_{n+1, 2n+1}-E_{n, n+1}) +
  \sum\limits_{1\leq i\leq n-1} i(2n+1-i)(E_{i, i+1}-E_{n+i+2, n+i+1})$,
\item"" $y = (E_{2n+1, n+1}-E_{n+1, n}) +
  \sum\limits_{1\leq i\leq n-1}(E_{i+1, i}-E_{n+i+1, n+i+2})$,
\item"" $z=t\bigr( (E_{2n-1, n+1}-E_{n+1, n-2})-(E_{2n+1, n-1}-E_{2n, n}) +
  \sum\limits_{1\leq i\leq n-3} (E_{i+3, i}-E_{n+i+1, n+i+4}) \bigl)$.
\endroster

Relations:
\roster
\item"{\bf 2.1}." $2 [z_1, \, z_2] - [z, \, z_3] = 144 t (2n^2+2n-9) z$,
\item"{\bf 2.2}." $9 [z_2, \, z_3] - 5 [z_1, \, z_4] =
    432 t (2n^2+2n-9)z_2 + 1728 t^2(n-1)(n+2)(2n-1)(2n+3) y$,
\item"{\bf 3.1}." $[z, \, [z, z_1]] = 0$,
\item"{\bf 3.2}." $7 [z_3, \, [z, z_1]] - 6 [z_2, \, [z, z_2]] =
  2880 t (n-3)(n+4)[z, z_1] $,
\item"$\pmb{\infty=n}$." $(\ad z_1)^{n-1}z = 0$.
\endroster

\noindent $\underline{\fsp(2n)\text{ for }n\geq 3}$. Generators:
\roster
\item"" $x=n^2E_{n, 2n} +
 \sum\limits_{1\leq i\leq n-1} i(2n-i)(E_{i, i+1}-E_{n+i+1, n+i})$,
\item"" $y=E_{2n, n} + \sum\limits_{1\leq i\leq n-1}(E_{i+1, i}-E_{n+i, n+i+1})$,
\item"" $z=t\biggl((E_{2n, n-2}+E_{2n-2, n}) - E_{2n-1, n-1} +
 \sum\limits_{1\leq i\leq n-3} (E_{i+3, i}-E_{n+i, n+i+3}) \biggr)$.
\endroster

Relations:
\roster
\item"{\bf 2.1}." $2 [z_1, \, z_2] - [z, \, z_3] = 72 t (4n^2-19) z$,
\item"{\bf 2.2}." $9 [z_2, \, z_3] - 5 [z_1, \, z_4] =
    216 t (4n^2-19)z_2 + 1728 t^2(n^2-1)(4n^2-9) y$,
\item"{\bf 3.1}." $[z, \, [z, z_1]] = 0$,
\item"{\bf 3.2}." $7 [z_3, \, [z, z_1] ] - 6 [z_2, \, [z, z_2]] =
     720 t (4n^2-49)[z, z_1] $,
\item"$\pmb{\infty=n}$." $(\ad z_1)^{n-1}z = 0$.
\endroster

\noindent $\underline{\fg_2}$. Generators:
\roster
\item"" $x=6X^+_1+10X^+_2, \qquad
y=X^-_1+X^-_2, \qquad
z={t\over 129600}[[X^-_1, X^-_2], [X^-_1, [X^-_1,  X^-_2]]]$.
\endroster

Relations:
\roster
\item"{\bf 2.1}." $[z, \, z_1] = 0$,
\item"{\bf 2.2}." $[z_1, \, z_2] = 0$,
\item"{\bf 2.3}." $[z_2, \, z_3] = -6tz$,
\item"{\bf 2.4}." $[z_3, \, z_4] = -8tz_2$,
\item"{\bf 2.5}." $[z_4, \, z_5] = -8tz_4+6t^2y$.
\endroster

\noindent $\underline{\ff_4}$.  Generators: \roster \item""
$x=16X^+_1+30X^+_2+42X^+_3+22X^+_4,\quad y=X^-_1+X^-_2+X^-_3+X^-_4$,
\item"" $z={t\over 907200}(2[[X^-_1, X^-_2], [X^-_3, [X^-_1,
X^-_2]]]+$\\
$2[[X^-_1, X^-_2], [X^-_4, [X^-_2, X^-_3]]]-[[X^-_3, X^-_4],
[X^-_2, [X^-_2, X^-_3]]])$.  \endroster

Relations:
\roster
\item"{\bf 2.1}." $[z, \, z_1] = 0$,
\item"{\bf 2.2}." $4 [z_2, \, z_3] - 9 [z_1, \, z_4] = 42 t z$,
\item"{\bf 2.3}." $5 [z_3, \, z_4] - 6 [z_2, \,
z_5] = 28 tz_2$,
\item"{\bf 2.4}." $13 [z_4, \, z_5] - 14 [z_3, \, z_6] =
    56 t z_4+ 306 t^2 y$.
\endroster

\noindent $\underline{\fe_6}$. Generators:
\roster
\item"" $x = 16X^+_1+30X^+_2+42X^+_3+30X^+_4+16X^+_5+22X^+_6$,
\item"" $y = X^-_1+X^-_2+X^-_3+X^-_4+X^-_5+X^-_6$,
\item"" $z = {t\over 8!} ([[X^-_1, X^-_2], [X^-_3, X^-_4]] -
 [[X^-_1, X^-_2], [X^-_3, X^-_6]] +$\\
 $ [[X^-_2, X^-_3], [X^-_4, X^-_5]]
 + [[X^-_3, X^-_6], [X^-_4, X^-_5]])$.
\endroster

Relations:
\roster
\item"{\bf 2.1}." $50[z_2, \, z_3] + 14 [z, \, z_5] - 35
  [z_1, \, z_4] = 0$,
\item"{\bf 2.2}." $20 [z_3, \, z_4] - 15 [z_2, \, z_5]
 + 7 [z_1, \, z_6] = 14 t^2 y$,
\item"{\bf 3.1}." $[z_1, \, [z, z_1]] = 0$,
\item"{\bf 3.2}." $[z_2, \, [z, z_1]] = 0$,
\item"{\bf 3.3}." $4 [z_3, \, [z, z_1] + 7 [z_1, \, [z_1, z_2]]
= 0$,
\item"{\bf 3.4}." $5 [z_3, \, [z, z_2]] + [z_4, \, [z, z_1] ]$,
\item"{\bf 3.5}." $8 [z_4, \, [z, z_2] ]
 + 5[z_3, \, [z_1, z_2]] = 0$,
\item"{\bf 3.6}." $3 [z_4, \, [z_1, z_2]]
 + 4 [z_4, \, [z, z_3]] = 0$,
\item"{\bf 3.7}." $ 51 [z_5, \, [z_1, z_2] ] +
 4 [z_5, [z, \, z_3] ] = -384 t^2 z$.
\endroster

\noindent $\underline{\fe_7}$.  Generators: \roster \item"" $x =
27X^+_1 + 52X^+_2 + 75 X^+_3 + 96 X^+_4 + 66 X^+_5 +34 X^+_6 + 49
X^+_7$, \item"" $y = X^-_1 + X^-_2 + X^-_3 + X^-_4 + X^-_5 + X^-_6 +
X^-_7$, \item"" $z = {7t\over 10!} ([[X^-_2, X^-_3], [X^-_7, [X^-_4,
X^-_5]]] + [[X^-_4, X^-_7], [X^-_5, [X^-_3, X^-_4]]] + $\\
$[[X^-_5,
X^-_6], [X^-_7, [X^-_3, X^-_4]]] + \brk 2 [[X^-_4, X^-_5], [X^-_3,
[X^-_1, X^-_2]]] + 2 [[X^-_5, X^-_6], [X^-_4, [X^-_2, X^-_3]]] -$\\
$ 3
[[X^-_4, X^-_7], [X^-_3, [X^-_1, X^-_2]]])$.  \endroster

Relations:
\roster
\item"{\bf 2.1}." $3 [z, z_5] -
    9 [z_1, z_4] + 14 [z_2, z_3] = - 2868 tz$,
\item"{\bf 2.2}." $18 [z_1, z_6] -
    50 [z_2, z_5] + 75 [z_3, z_4] = - 9560 tz_2$,
\item"{\bf 2.3}." $14 [z_2, z_7] - 35 [z_3, z_6]
    + 50 [z_4, z_5] = - 4780 tz_4 + 49335 t^2y$;
\item"{\bf 3.1}." $[z, [z, z_1]] = 0$,
\item"{\bf 3.2}." $9 [z_1, [z, z_2]] - 4 [z_2, [z, z_1]] = 0$,
\item"{\bf 3.3}." $330 [z_2, [z, z_2]] - 425 [z_3, [z, z_1]]
     - 1458 [z_1, [z_1, z_2]] = 0$,
\item"{\bf 3.4}." $665 [z_3, [z, z_2]] -
    640 [z_4, [z, z_1]] -
    1134 [z_2, [z_1, z_2]] = 0$,
\item"{\bf 3.5}." $5485 [z_3, [z, z_3]] - 3910 [z_4, [z, z_2]] -
    3182 [z_3, [z_1, z_2]] =
       2527815 t[z, z_1]$,
\item"{\bf 3.6}." $825 [z_4, [z, z_3]] - 598 [z_5, [z, z_2]]
    - 876 [z_4, [z_1, z_2]] = 338422 [z, z_2]$,
\item"{\bf 3.7}." $1525 [z_5, [z, z_3]]
    - 7524 [z_5, [z_1, z_2]] +
2415 [z_4, [z_1, z_3]]= 1106875 t[z, z_3] + 2734746 t [z_1, z_2]\,$
\item"{\bf 3.8}." $25250 [z_6, [z, z_4]]
    - 94920 [z_6, [z_1, z_3]] +
    44252 [z_5, [z_1, z_4]] =\brk
- 1305480 t[z, z_5] + 41398712 t[z_1, z_4] - 1117925005 t^2z$;
\item"{\bf 4.1}." $12[[z, z_2], [z_1, z_2]] -
    5[[z, z_2], [z, z_3]] = 0$,
\item"{$\pmb{\infty=5}$}." $[[z, z_2], [z_1, [z, z_1]]] = 0$.
\endroster

\noindent $\underline{\fe_8}$. Generators:
\roster
\item"" $x = 58X^+_1 + 114X^+_2 + 168X^+_3 + 220X^+_4 + 270X^+_5 + 182X^+_6 +
    92X^+_7 + 136X^+_8$,
\item"" $y = X^-_1 + X^-_2 + X^-_3 + X^-_4 + X^-_5 + X^-_6 + X^-_7 + X^-_8$,
\item"" $z = {t\over13!}\bigl([[X^-_7, [X^-_5, X^-_6]], [[X^-_3, X^-_4], [X^-_5,
X^-_8]]] +
[[X^-_8, [X^-_4, X^-_5]], [[X^-_3, X^-_4], [X^-_5, X^-_6]]] +\brk
[[X^-_8, [X^-_5, X^-_6]], [[X^-_1, X^-_2], [X^-_3, X^-_4]]] +
[[X^-_8, [X^-_5, X^-_6]], [[X^-_2, X^-_3], [X^-_4, X^-_5]]] +\brk
[[X^-_8, [X^-_5, X^-_6]], [[X^-_4, X^-_5], [X^-_6, X^-_7]]] +\brk
2 [[X^-_4, [X^-_2, X^-_3]], [[X^-_5, X^-_8], [X^-_6, X^-_7]]] -
3 [[X^-_7, [X^-_5, X^-_6]], [[X^-_1, X^-_2], [X^-_3, X^-_4]]]\bigr)$.
\endroster

Relations:
\roster
\item"{\bf 2.1}." $91 [z, z_5] -
    325 [z_1, z_4] + 550 [z_2, z_3] = 0$,
\item"{\bf 2.2}." $13 [z_1, z_6] -
    45 [z_2, z_5] + 75 [z_3, z_4] =
      - 268814 tz$,
\item"{\bf 2.3}." $33 [z_2, z_7] -
    11 [z_3, z_6] + 180 [z_4, z_5] =
      - 682374 tz_2$,
\item"{\bf 2.4}." $11 [z_3, z_8] -
    35 [z_4, z_7] + 56 [z_5, z_6] =
      - 186102t z_4$,
\item"{\bf 2.5}." $3 [z_4, z_9] -
    9 [z_5, z_8] + 14 [z_6, z_7]=
      - 41356 tz_6
      + 2686866 t^2y$;
\item"{\bf 3.1}." $[z, [z, z_1]] = 0$,
\item"{\bf 3.2}." $13 [z_1, [z, z_2]] - 6 [z_2, [z, z_1]] = 0$,
\item"{\bf 3.3}." $542 [z_2, [z, z_2]]
    - 639 [z_3,  [z, z_1]] - 2236 [z_1, [z_1, z_2]] = 0$,
\item"{\bf 3.4}." $1067 [z_3, [z, z_2]] -
    950 [z_4,  [z, z_1]] -
    1892 [z_2, [z_1, z_2]] = 0$,
\item"{\bf 3.5}." $7255 [z_3, [z, z_3]] -
    4995 [z_4,  [z, z_2]] - 4527 [z_3, [z_1, z_2]] = 0$,
\item"{\bf 3.6}." $105460 [z_4, [z, z_3]]
    - 69597 [z_5, [z, z_2]]
    - 119430 [z_4, [z_1, z_2]] = 0$,
\item"{\bf 3.7}." $844277 [z_5,  [z, z_3]]
    + 1556775 [z_4, [z_1, z_3]]
    - 4442058 [z_5, [z_1, z_2]] =
- 17362538193 t[z, z_1]$,
\item"{\bf 3.8}." $334453 [z_6, [z, z_4]]
    + 746586 [z_5, [z_1, z_4]]
    - 1414050 [z_6, [z_1, z_3]] =\brk
1120518212 t[z, z_3] +
      3082429152 t[z_1, z_2]$,
\item"{$\pmb{\infty=4}$ }." $[[z, z_1], [z, z_2]] = 0$.
\endroster
\subhead   Table 3.2. $\fsl(\lambda)$ and $\fo{/}\fsp(\lambda)$\endsubhead

\noindent $\fsl(\lambda)$.
Generators:
\roster
\item"" $x = u^2{d\over d u} - (\lambda-1) u, \qquad
y = - {d\over d u}, \qquad
z = t{d^2 \over d u^2}$.
\endroster

Relations:
\roster
\item"{\bf2.1}." $3[z_1, \, z_2] - 2[z, \, z_3] = 24 t^2(\lambda^2-4)
y$,
\item"{\bf3.1}." $[z, \, [z, z_1] = 0$,
\item"{\bf3.2}." $4 [z_3, \, [z, z_1] ] - 3 [z_2, \, [z, z_2]]
    = 576 t^2(\lambda^2-9)z$.
\endroster

\noindent $\fo{/}\fsp(\lambda)$. Generators:
\roster
\item"" $x = u^2{d\over d u} - (\lambda-1) u, \qquad
y = - {d\over d u}, \qquad
z = t{d^3 \over d u^3}$.
\endroster

Relations:
\roster
\item"{\bf2.1}." $2 [z_1, \, z_2] - [z, \, z_3] = 72 t (\lambda^2-19)
z$,
\item"{\bf2.2}." $9 [z_2, \, z_3] - 5 [z_1, \, z_4] =
    216 t (\lambda^2-19)z_2 - 432 t^2(\lambda^2-4)(\lambda^2-9) y$,
\item"{\bf3.1}." $[z, \, [z, z_1] = 0$,
\item"{\bf3.2}." $7 [z_3, \, [z, z_1]] - 6 [z_2, \, [z, z_2]] =
  720 t (\lambda^2-49)[z, z_1] $ .
\endroster

\subhead Table 3.3. $\fsl(*)$ and $\fo{/}\fsp(*)$\endsubhead

\noindent $\underline{\fsl(*)}$.

\roster
\item"{\bf2.1}." $3[z_1, z_2] - 2[z, z_3] = 24 t^2 y$,
\item"{\bf3.1}." $[z, \, [z, z_1] = 0$,
\item"{\bf3.2}." $4 [[z, z_1], z_3]]] + 3 [z_2, [z, z_2]]
    = -576 t^2 z$.
\endroster

\noindent $\underline{\fo{/}\fsp(*)}$.

\roster
\item"{\bf2.1}." $2 [z_1, z_2] - [z, z_3] = 72 t z$,
\item"{\bf2.2}." $9 [z_2, z_3] - 5 [z_1, z_4] =
    216 t z_2 - 432 t^2 y$,
\item"{\bf3.1}." $[z, \, [z, z_1] = 0$,
\item"{\bf3.2}." $7 [[z, z_1], z_3] + 6 [z_2, [z, z_2]] =
    - 720 t [z, z_1] $.
\endroster

\head \S 4. Remarks\endhead

\subhead 4.1.  On proof\endsubhead For the exceptional cases the
proof is direct: the quotient of the free Lie algebra generated by
$x, y$ and $z$ modulo our relations is the needed finite
dimensional one. For rank $\fg\leq 10$ we similarly computed
relations for $\fg=\fsl$, $\fo$ and $\fsp$; as Post pointed out,
together with the result of \cite{PH} on deformation (cf.  2.7)
this completes the proof.

Our theorems elucidate Proposition 2 of \cite{F}; we just wrote
relations explicitly.  Feigin claimed \cite{F} that, for
$\fsl(\lambda)$, the relations of type 3 follow from the
decomposition of
$$
L^{k_1}\wedge L^{k_2}\subset L^{k_1}\wedge
L^{k_1}\wedge L^{k_1}.
$$
We verified that this is so not only in
Feigin's case but for all the above-considered algebras except
$\fe_6$, $\fe_7$ and $\fe_8$: for them one should consider the
whole $L^{k_1}\wedge L^{k_1}\wedge L^{k_1}$.

\proclaim {4.2. Proposition}.
(a) For a principal embedding $\fsl(2)\longrightarrow\fg$, where $\fg=\fo(2n+1)$,
$\fsp(2n)$ or $\fo{/}\fsp(\lambda)$, where $\lambda\in {\Bbb C}{\Bbb P}^1$, there exists an
embedding of
$\sigma: \fg\longrightarrow\fsl(k)$ for an appropriate $k\in {\Bbb C}{\Bbb P}^1$ such that the through
map is principal.

(b) There is no such $\sigma$ for the exceptional Lie algebras or
$\fo(2n)$.
\endproclaim

\subhead 4.3. How to present$\fo(2n)$?\endsubhead
Select $z$ as in sec. 1.1. Clearly, the form of $z$ (hence, relations of type 1) and the
number of relations of type 3 depend on $n$; this was not the case for the algebras
considered above. Besides, the relations are not as neat as for the above algebras.

\noindent $\underline{\fo(2n)}$. Generators:
\roster
\item"" $x = {n(n-1)\over2}
 (E_{n-1,n}-g_{2n,2n-1} + E_{n-1,2n}-E_{n,2n-1})
 +$\\
 $ \sum\limits_{1\leq i\leq n-2}\left(i(2n-1-i)(E_{i,i+1}-E_{n+i+1,n+i})\right)$,
\item"" $y = (E_{2n,n-1}-E_{2n-1,n}) +
  \sum\limits_{1\leq i\leq n-1}(E_{i+1,i}-E_{n+i,n+i+1})$,
\item"" $z = {1\over(2n-2)!}
  \left((E_{n,1}-E_{n+1,2n}) + (E_{n+1,n} - E_{2n,1})\right)$.
\endroster

We can not
write the relations in full generality; for small values of $n$ they are:

\noindent $\underline{n=4}$.
$$
\matrix
&\pmb{2.1}.&3 [z, z_5]
        - 5 [z_1, z_4] + 6 [z_2, z_3]= \frac{1}{2} y;&&\pmb{3.5}.&[z_3, [z, z_1]] = 0,\cr
&\pmb{3.1}.&[z, [z, z_1]] = 0,&&\pmb{3.6}.&[z_3, [z, z_2]] = 0,\cr
&\pmb{3.2}.&[z_1,[z, z_1]] = 0,&&\pmb{3.7}.&[z_4, [z, z_2]] = z,\cr
&\pmb{3.3}.&[z_2, [z, z_1]] = 0,& &\pmb{3.8}.&[z_4, [z_1, z_2]] = z_1,\cr
&\pmb{3.4}.&[z_1, [z_1, z_2]] = 0,& &\pmb{3.9}.&[z_5, [z_1, z_2]] = z_2.\cr
\endmatrix
$$

\noindent $\underline{n = 5}$. There are 17 relations of type 3;
the relation of type 2 is:
\roster
\item"{\bf 2.1}. " $- 4 [z, z_7] + 7 [z_1, z_6] - 9 [z_2, z_5] + 10 [z_3, z_4] = \frac{1}{2}
y$,
\endroster

\noindent $\underline{n = 6}$.  The relation of type 2 is still more
involved and there are 27 relations of type 3.

We should, perhaps, have taken the generators as for $\fo(2n-1)$ and
add a generator from $L^{2n-2}$.  We have no guiding idea; to try at
random is frustrating.

\subhead 4.4.  How to realize $\fsl(*)$ and
$\fo{/}\fsp(*)$?\endsubhead We do not know how to answer this question
and while this is a research problem we can not express the Jacobson
generators in a form as suggestive as for $\lambda\in{\Bbb C}$.

\subhead 4.5.  A relation with integrable differential
systems\endsubhead The Drinfeld--Sokolov's construction, as well as
its generalization to $\fsl(\lambda)$ and $\fo{/}\fsp(\lambda)$
(\cite{DS}, \cite{KM}), hinges on a certain element that can be
identified with the image of $X^+\in \fsl(2)$ under the principal
embedding.  One might think that only this image is important but the
image of the whole $\fsl(2)$ is recovered from the image of $X^+$
whereas to work with $\fsl(2)$ is easier than with a nilpotent element
--- the image of $X^+$.

\enddocument